\documentclass[a4paper,12pt,leqno]{article}

\usepackage{latexsym}
\usepackage{bm,amsfonts,amsmath,amssymb} 
\usepackage[utf8]{inputenc}
\usepackage{abstract}
\usepackage[T1]{fontenc}
\usepackage{graphicx}
\usepackage{verbatim}
\usepackage{ulem}

\usepackage{epsfig}
\usepackage{color}

\usepackage[perpage]{footmisc}
\usepackage{fancybox}
\usepackage{dsfont}

\begin{document}

\newcommand{\R}{\mathbb{R}}
\newcommand{\C}{\mathbb{C}}
\newcommand{\N}{\mathbb{N}}
\newcommand{\Z}{\mathbb{Z}}
\newcommand{\Q}{\mathbb{Q}}
\newcommand{\mj}{\mathcal}

\newcommand{\su}{\mathrm{supp}}
\newcommand{\la}{\lambda}
\newcommand{\fcv}{\rightharpoonup}
\newcommand{\bg}{\begin}
\newcommand{\ds}{\displaystyle}
\newcommand{\Om}{\Omega}
\newcommand{\eps}{\epsilon}
\newcommand{\Sp}{\mathbb{S}}
\newcommand{\inj}{\hookrightarrow}
\newcommand{\n}{\textbf{n}}
\newcommand{\bv}{\textbf{b}}
\newcommand{\h}{\textbf{h}}
\newcommand{\A}{\textbf{A}}
\newcommand{\F}{\textbf{F}}
\newcommand{\B}{\textbf{B}}
\newcommand{\dr}{\partial}
\newcommand{\vv}{\textbf{v}}
\newcommand{\uu}{\textbf{u}}
\newcommand{\tr}{\mathrm{Tr}}
\newcommand{\conj}{\overline}
\newcommand{\intn}{\int_{\Om}\!\!\!\!\!\!\!-}
\newcommand{\dive}{\mathrm{div}}

\newtheorem{lem}{Lemma}[section]
\newtheorem{theo}[lem]{Theorem}
\newtheorem{prop}[lem]{Proposition}
\newtheorem{sch}[lem]{Scholie}
\newtheorem{cor}[lem]{Corollary}
\newtheorem{conjec}[lem]{Conjecture}

\def\got#1{{\bm{\mathfrak{#1}}}}

\newenvironment{preuve}
{\noindent{\textbf{Proof.}}\\\rm\noindent}
{\bg{flushright}\tiny $\blacksquare$\end{flushright}}

\newenvironment{rem}
{\noindent\addtocounter{lem}{1}
{\textbf{Remark \thelem.}}\\\noindent\rm}
{\bg{flushright}\tiny $\blacksquare$\end{flushright}}

\renewcommand{\theequation}{\thesection.\arabic{equation}}
\renewcommand{\thefigure}{\thesection.\arabic{equation}}

\title{Sharp asymptotics for the Neumann Laplacian with variable magnetic field : case of dimension 2}
\author{Nicolas Raymond\footnote{Université Paris-Sud 11\newline Bâtiment 425, Laboratoire de Mathématiques\newline 91405 Orsay Cedex\newline e-mail : nicolas.raymond@math.u-psud.fr}}
\maketitle{}

\bg{abstract}
The aim of this paper is to establish estimates of the lowest eigenvalue of the Neumann realization of $(i\nabla+B\A)^2$ on an open bounded subset of $\R^2$ $\Om$ with smooth boundary as $B$ tends to infinity. We introduce a "magnetic" curvature mixing the curvature of $\dr\Om$ and the normal derivative of the magnetic field and obtain an estimate analogous with the one of constant case. Actually, we give a precise estimate of the lowest eigenvalue in the case where the restriction of magnetic field to the boundary admits a unique minimum which is non degenerate. We also give an estimate of the third critical field in Ginzburg-Landau theory in the variable magnetic field case.
\end{abstract}

\section{Introduction and statement of main results}
Let $\Om$ be an open bounded subset of $\R^2$ with smooth boundary and $\A~\in~\mj{C}^{\infty}(\conj{\Om},\R^2)$. We let :
$$\beta=\nabla\times\A$$ 
and for $B>0$ and $u\in H^1(\Om)$ :
$$q^N_{B\A,\Om}(u)=\int_{\Om}|(i\nabla+B\A)u|^2 dx$$
and we consider the associated selfadjoint operator, i.e the Neumann realization of $(i\nabla+B\A)^2$ on $\Om$. We denote by $\lambda_1(B\A)$ the lowest eigenvalue of this operator. By the minimax principle, we have :
$$\lambda_1(B\A)=\inf_{u\in H^1(\Om)}\frac{q^N_{B\A,\Om}(u)}{\|u\|^2}.$$
We first recall some properties of the harmonic oscillator on a half axis (see \cite{DauHel,HelMo3}).
\paragraph{Harmonic oscillator on a half axis}~\\
For $\xi\in\R$, we consider the Neumann realization $\got{h}^{N,\xi}$ in $L^2(\R_+)$ associated with the operator \bg{equation}\label{oh}
-\frac{d^2}{dt^2}+(t+\xi)^2,\quad\mj{D}(\got{h}^{N,\xi})=\{u\in B^2(\R_+) : u'(0)=0\}.
\end{equation}
One knows that it has compact resolvent and its lowest eigenvalue is denoted $\mu(\xi)$ ; the associated $L^2$-normalized and positive eigenstate is denoted by $u_{\xi}~=~u(\cdot,\xi)$ and is in the Schwartz class. The function $\xi\mapsto\mu(\xi)$ admits a unique minimum in $\xi=\xi_0$ and we let : 
\bg{equation}
\Theta_0=\mu(\xi_0),
\end{equation}
\bg{equation}
C_1=\frac{u^2_{\xi_0}(0)}{3}.
\end{equation}
Let us also recall identities established by \cite[p. 1283-1284]{BeSt}. For $k\in\N^*$, we denote by $M_k$ :
$$M_k=\int_{t>0}(t+\xi_0)^k |u_{\xi_0}(t)|^2 dt.$$
\bg{align}\label{formulasBS}
M_0=1,\quad & M_1=0,\quad & M_2=\frac{\Theta_0}{2},\quad& M_3=\frac{C_1}{2}\quad&\mathrm{and}\quad\frac{\mu''(\xi_0)}{2}=3C_1\sqrt{\Theta_0}.
\end{align}
Let us state a result in the case where $\beta$ is constant :
\bg{theo}
Assuming that $\beta=1$, we have the estimate :
$$\lambda_1(B\A)=\Theta_0 B-C_1\kappa_{max}\sqrt{B}+O(B^{1/3}),$$
where $$\kappa_{max}=\max\{k(s), s\in\dr\Om\}$$
and $k(s)$ denotes the curvature of the boundary at the point $s$.
Moreover, the grounstate decays exponentially away from the points of maximal curvature. 
\end{theo}
\bg{rem}
This result was first announced by a formal analysis in \cite{BeSt} and rigorously proved in the case of the disk (see \cite{BPT}). Let us 
also mention that in \cite{Lupan}, an estimate at the first order was rigorously proved (see also \cite{Lupan4} for the problem in $\R^2$ and $\R^2_+$). For higher order expansion in the case of constant magnetic field, one can finally mention \cite{PFS,HelMo3,FouHel3,FouHel2}.
\end{rem}
Our aim is to obtain a similar result when the the magnetic field is not constant.
We will assume that $\beta>0$ on $\conj{\Om}$.  
We introduce : 
\bg{eqnarray}\label{bbprime}
b=\inf_{\Om}\beta & \mathrm{and} & b'=\inf_{\dr\Om}\beta,
\end{eqnarray}
and we assume : 
\bg{equation}\label{theta0b'}
\Theta_0 b'<b.
\end{equation}
\paragraph{Estimate for the variable magnetic field}~\\
Let us state a first (rough) estimate concerning the first eigenvalue :
\bg{theo}\label{roughestimate}
Assuming that $\beta_{|\dr\Om}$ admits a unique and non degenerate minimum, we have :
$$\lambda_1(B\A)=\Theta_0 b'B+O(B^{1/2}).$$
\end{theo}
\bg{rem}
The first term was obtained by many authors (cf. \cite{Lupan,HelMo3}) with a worse remainder estimate. Our assumption of non-degeneracy permits to find the optimal remainder $O(B^{1/2})$ (the improvement occurs for the lower bound) which is crucial to establish tangential Agmon estimates (see Section 4).
\end{rem}
Let us also state a tangential localization result of the first eigenfunctions :
\bg{prop}[Tangential Agmon's estimates for $u_B$]\label{tagm1}
Let $u_B$ be an eigenfunction associated with the lowest eigenvalue of the Neumann realization of $(i\nabla+B\A)^2$.
We have the control :
$$\int\exp(\alpha_1\chi(t(x)) d(s(x)) B^{1/2})\{|u_B|^2+B^{-1}|(i\nabla+B\A)u_B|^2\}dx\leq C\|u_B\|^2,$$
where $\chi$ is a smooth cutoff function in a neighborhood of the boundary, $t(x)=d(x,\dr\Om)$, $s(x)$ the curvilinear coordinate on the boundary and where $d$ is the Agmon distance to the minimum of $\beta$ defined in Section 4.
\end{prop}
\bg{rem}
This estimate improves the localization found in \cite{HelMo3} by specifying the behaviour of $u_B$ near the minimum of $\beta$.
In Section 4 we also get tangential Agmon estimates for $D_s u_B$. All these localizations properties are essential to obtain the second correction term of Theorem \ref{roughestimate}.
\end{rem}
\bg{theo}\label{preciseestimate}
Assuming that $\beta_{|\dr\Om}$ admits a unique and non degenerate minimum in $x_0$, we have :
$$\lambda_1(B\A)=\Theta_0b' B+\Theta_{1/2}b'^{1/2} B^{1/2}+O(B^{2/5}),$$
where $$\Theta_{1/2}=\Theta_{1/2}(x_0)=-\kappa(x_0)C_1+\left(\frac{C_1}{2}-\Theta_0\xi_0\right)\frac{1}{b'}\frac{\dr\beta}{\dr t}(x_0)+\Theta_0^{3/4}\left(\frac{3C_1}{2b'}\frac{\dr^2\beta}{\dr s^2}(x_0)\right)^{1/2}.$$
\end{theo}
\bg{rem}
\vspace*{-0.5cm}
\bg{enumerate}
\item When $\beta_{|\dr\Om}$ admits a finite set $\mj{M}$ of non degenerate minima, we have the same expansion by replacing $\Theta_{1/2}$ by $\ds{\min_{x\in\mj{M}}\Theta_{1/2}(x)}$.
\item Without assuming the non degeneracy of the minima, we believe that the conclusion of Theorem \ref{preciseestimate} is true by replacing $\Theta_{1/2}$ by $\ds{\min_{x\in\mj{M}}\Theta_{1/2}(x)}$.
\item The optimal remainder is certainly $O(B^{1/4})$ as suggested by the upper bound.
\item The computations for the upper bound suggest the following expansion of the $n$-th eigenvalue :
$$\la^n(B\A)=\Theta_0b'B+\Theta_{1/2}^n B^{1/2}+O(B^{1/4}).$$
where :
$$\Theta_{1/2}^n=-\kappa(x_0)C_1+\left(\frac{C_1}{2}-\Theta_0\xi_0\right)\frac{1}{b'}\frac{\dr\beta}{\dr t}(x_0)+(2n-1)\Theta_0^{3/4}\left(\frac{3C_1}{2b'}\frac{\dr^2\beta}{\dr s^2}(x_0)\right)^{1/2}.$$
\item In the variable case, the localization due to the curvature doesn't play a role anymore ; the effect of the curvature is small compared to the variation of the magnetic field.
\item This expansion with two terms of the first eigenvalue could be generalized at any order under the previous assumptions (unique and non degenerate minimum of $\beta_{|\dr\Om}$) by using a Grushin approach (see \cite{FouHel3}).
\item The case where the magnetic field (non degenerately) vanishes in $\Om$ was treated in \cite{KwPan}. Moreover, the case where it non degenerately vanishes on the boundary remains open and should be an interesting problem. 
\item Theorems \ref{roughestimate} and \ref{preciseestimate} are also sensible under the hypothesis of regularity of the domain. When the domain has corners (see \cite[Theorem 1.2]{Bona}) and with a variable magnetic field, the ground state is not necessarily localized near the points of the boundary where the magnetic field is minimum.
\item The asymptotic behaviour in Theorems \ref{roughestimate} and \ref{preciseestimate} is strongly dependent on the Neumann boundary condition we impose, as one can see in \cite{Kachmar,Kachmar2,Kachmar3}. In particular, in certain cases, the localization is no more determined by the minimal points of $\beta$.
\end{enumerate}
\end{rem}
\paragraph{Constant magnetic field on the boundary}~\\
In \cite{Aramaki,Aramaki2}, the case  of the constant magnetic field on the boundary is treated. Nevertheless, this case is studied under a non degeneracy condition : it is assumed that the curvature of the boundary $\kappa$ admits a unique maximum at $x=x_0$ and that the normal derivative $\ds{\frac{\dr\beta}{\dr t}}$ admits a unique minimum at $x=x_0$ ; moreover, the minimum of $\ds{\frac{\dr\beta}{\dr t}-b'\kappa}$ has to be non degenerate. 
Here, we improve his result by using more generic assumptions ; in particular, we will see that the quantity to maximize is the "magnetic curvature" defined by :
$$\tilde{\kappa}(x)=C_1\kappa(x)+\left(\Theta_0\xi_0-\frac{C_1}{2}\right)\frac{1}{b'}\frac{\dr\beta}{\dr t}(x).$$
More precisely, our result is the following :
\bg{theo}[Upper bound : constant magnetic field on $\dr\Om$]\label{constantfield}
When the magnetic field is constant on the boundary, we have the upper bound : 
$$\la^1(B\A)\leq\Theta_0 b' B-\max_{x\in\dr\Om}\left\{C_1\kappa(x)-\left(\frac{C_1}{2}-\Theta_0\xi_0\right)\frac{1}{b'}\frac{\dr\beta}{\dr t}(x)\right\}b'^{1/2}B^{1/2}+O(B^{1/3}),$$
where $\kappa(x) denotes$ the curvature of the boundary at $x$.
\end{theo}
\bg{rem}
\bg{enumerate}
\vspace*{-0.5cm}
\item The corresponding lower bound could certainly be obtained by the techniques of \cite{FouHel2}.
\item Assuming the existence of a unique and non degenerate maximum of the magnetic curvature $\tilde{\kappa}$, one could surely give an asymptotics at any order of $\la^1(B\A)$ and localization properties as for the constant magnetic field case (see \cite{FouHel3}) which would improve the hypothesis of Aramaki.
\end{enumerate}
\end{rem}
\paragraph{Organization of the paper}~\\
In Section 2 and 3, we will prove the Theorem \ref{roughestimate} and give the upper bound of Theorem \ref{preciseestimate} and of Theorem \ref{constantfield}. Then, we will see, in Section 4, that this first rough estimate gives information on the localization of the groundstates on the boundary near the mimimum of the magnetic field. In Section 5, we prove the lower bound of Theorem \ref{preciseestimate} thanks to a reduction to a degenerate case studied by S. Fournais and B. Helffer.
Finally, we apply the previous results to give an estimate of the third critical field in Ginzburg-Landau theory.
\section{A rough lower bound}
In order to get the lower bound in Theorem \ref{roughestimate}, we use a localization technique permitting the reduction to easier models.
\subsection{Partition of unity}
For each $0<\rho<\frac{1}{2}$, $B>0$, $\eps>0$ and $C_0>0$, we consider a partition of unity (cf. \cite{HelMo2}) for which there exists $C=C(\Om,\beta,\eps,C_0)>0$ such that :
\bg{eqnarray}\label{partition}
\sum_j |\chi_j^B|^2=1 \mbox{ on }\Om\,;\\
\sum_j |\nabla\chi_j^B|^2\leq CB^{2\rho} \mbox{ on }\Om.\\\nonumber
\end{eqnarray}
Each $\chi_j^B$ is a $\mj{C}^{\infty}$-cutoff function supported in $D_j\cap\conj{\Om}$. Moreover, we may assume that there exists a ball $D_j=D_{j_{min}}$ whose center is the minimum of $\beta$ on the boundary and $C_0 B^{-\rho}$ for radius. We may also assume that the balls which intersect the boundary have their centers on the boundary and that those one admit $\eps B^{-\rho}$ for radius. The radius of all the other balls is assumed to be $B^{-\rho}$.  
We will choose $\rho$, $\eps$ and $C_0$ later for optimizing the error.
We will use the following localization IMS formula (cf. \cite{Cycon}) :
\bg{lem}
\bg{eqnarray}\label{IMS}
q_{B\A}(u)=\sum_j q_{B\A}(\chi_j^B u)-\sum_j \||\nabla\chi_j^B|u\|^2,\qquad\forall u\in H^1(\Om).
\end{eqnarray}
\end{lem}
So, in order to minimize $q_{B\A}(u)$, we are reduced to the minimization of $q_{B\A}(v)$, with $v$ supported in some $D_j$.
\subsection{Estimates for the lower bound}
\subsubsection{Study inside $\Om$}
Let $j$ such that $D_j$ does not intersect the boundary.
It is well known that : $$q_{B\A}(\chi_j^B u)\geq B\int_{\Om}\beta(x)|\chi_j^B u|^2 dx\geq bB\int_{\Om}|\chi_j^B u|^2 dx.$$
Having in mind (\ref{theta0b'}), these terms will not play a role in the computation of the asymptotics.
\subsubsection{Study at the boundary}
In the next paragraph, we introduce boundary coordinates.
\paragraph{Boundary coordinates}~\\
We choose a parametrization of the boundary : 
$$\gamma : \R/(|\dr\Om|\Z)\to\dr\Om.$$
Let $\nu(s)$ be the unit vector normal to the boundary, pointing inward at the point $\gamma(s)$. We choose the orientation of the parametrization $\gamma$ to be counter-clockwise, so
$$\det(\gamma'(s),\nu(s))=1.$$
The curvature $k(s)$ at the point $\gamma(s)$ is given in this parametrization by :
$$\gamma''(s)=k(s)\nu(s).$$ 
The map $\Phi$ defined by :
\bg{eqnarray*}
\Phi : \R/(|\dr\Om|\Z)\times]0,t_0[\to\Om\\
(s,t)\mapsto\gamma(s)+t\nu(s),
\end{eqnarray*}
is clearly a diffeomorphism, when $t_0$ is sufficently small, with image
$$\Phi(\R/(|\dr\Om|\Z)\times]0,t_0[)=\{x\in\Om|d(x,\dr\Om)<t_0\}=\Om_{t_0}.$$ 
We let :
$$\tilde{A}_1(s,t)=(1-tk(s))\A(\Phi(s,t))\cdot\gamma'(s),\quad\tilde{A}_2(s,t)=\A(\Phi(s,t))\cdot\nu(s),$$
$$\tilde{\beta}(s,t)=\beta(\Phi(s,t)),$$
and we get :
$$\dr_s\tilde{A}_2-\dr_t\tilde{A}_1=(1-tk(s))\tilde{\beta}(s,t).$$
Let $j$ such that $B_j$ intersect the boundary ; we have, with $v_j=\chi_j^B u$ and $\tilde{v}_j=v_j\circ\Phi$ :
$$q_{B\A}(v_j)=\int(1-tk(s))|(i\dr_t+B\tilde{A}_2)\tilde{v}_j|^2+(1-tk(s))^{-1}|(i\dr_s+B\tilde{A}_1)\tilde{v}_j|^2 ds dt.$$
\paragraph{Approximation by a constant magnetic field on a domain with constant curvature}~\\
Locally, we can choose a gauge such that 
\bg{eqnarray*}
\tilde{A}_1(s,t)=\int_0^t(1-t'k(s))\tilde{\beta}(s,t')dt',&\tilde{A}_2=0.
\end{eqnarray*}
We assume that the center of the ball $D_j$ has the coordinates $(s_j,0)$ and that the coordinates of the minimum are $(0,0)$. We let :
$$k_j=k(s_j),\quad \tilde{\beta}(s_j,0)=\tilde{\beta}_j\quad \mbox{and}\quad \Delta k_j(s)=k(s)-k_j.$$
We have : 
\bg{align}\label{approxbeta}
(1-tk(s))\tilde{\beta}(s,t)=(1-tk_j)\tilde{\beta}_j-t\Delta k_j(s)\tilde{\beta}(s,t)+(1-tk_j)(\tilde{\beta}(s,t)-\tilde{\beta}_j).
\end{align}
We write :
\bg{align}\label{approxA}
\tilde{A}_1(s,t)=\conj{A}_{1,j}(s,t)+R_j(s,t),
\end{align}
with 
\bg{align}\label{Abarre}
\conj{A}_{1,j}(s,t)=(t-k_j\frac{t^2}{2})\tilde{\beta}_j.
\end{align}
\paragraph{Control of the remainders}~\\
Therefore, we are reduced to compare $q_{B\A}$ with the quadratic form associated with the Neumann problem on a domain with constant curvature (see \cite{BPT,FouHel2,FouHel3,HelMo3}).
For all $\la>0$, we get the inequality (with the Cauchy-Schwarz inequality):
\bg{eqnarray*}
q_{B\A}(v_j)&\geq&(1-\la)\int (1-tk_j)|\dr_t\tilde{v}_j|^2+(1-tk_j)^{-1}|(i\dr_s+B\conj{A}_{1,j})\tilde{v}_j|^2 ds dt\\
    &    &-C\int\Delta k_j(s)t (|\dr_t\tilde{v}_j|^2+|(i\dr_s+B\tilde{A}_1)\tilde{v}_j|^2) dsdt\\
    &    &-\frac{B^2}{\la}\int|R_j(s,t)\tilde{v}_j|^2 dsdt.
\end{eqnarray*}
We apply the result of the constant magnetic field on a domain with constant curvature to get the existence of $C>0$ such that for all $j$ such that ${D_j\cap\dr\Om\neq\emptyset}$ (cf. \cite[Theorem 6.1]{BPT})~:
\bg{equation}\label{diskcase}
\int (1-tk_j)|\dr_t\tilde{v}_j|^2+(1-tk_j)^{-1}|(i\dr_s+B\conj{A}_{1,j})\tilde{v}_j|^2 ds dt\geq (\Theta_0 \tilde{\beta}_jB-C_1k_j B^{1/2}-C)\|\tilde{v}_j\|^2.
\end{equation}
In order to control the remainders, we recall the Agmon estimates (cf. \cite{Agmon, HelMo3, FouHel2, FouHel3}) :
\bg{prop}[Normal Agmon's estimates]\label{normalagmon}
Let $u_B$ be an eigenfunction associated with the lowest eigenvalue of the Neumann realization of $(i\nabla+B\A)^2$.
We have the control the momenta of order $n$ in the normal variable $t$ :
$$\int t(x)^n\{|u_B|^2+B^{-1}|(i\nabla+B\A)u_B|^2\}dx\leq C_n B^{-\frac{n}{2}}\|u_B\|^2.$$
\end{prop}
We choose $\rho=\frac{1}{4}$ (see Figure \ref{figurepartition}) and notice that $|\Delta_jk(s)|=O(B^{-1/4})$ (uniformly in $j$). 
\bg{figure}[h]
\bg{center}
\scalebox{0.7}
{
\begin{picture}(0,0)%
\includegraphics{figure.pstex}%
\end{picture}%
\setlength{\unitlength}{3947sp}%
\begingroup\makeatletter\ifx\SetFigFont\undefined%
\gdef\SetFigFont#1#2#3#4#5{%
  \reset@font\fontsize{#1}{#2pt}%
  \fontfamily{#3}\fontseries{#4}\fontshape{#5}%
  \selectfont}%
\fi\endgroup%
\begin{picture}(9174,4505)(589,-5309)
\put(1651,-3436){\makebox(0,0)[lb]{\smash{{\SetFigFont{12}{14.4}{\rmdefault}{\mddefault}{\updefault}{\color[rgb]{0,0,0}$\epsilon B^{-1/4}$}%
}}}}
\put(9151,-1111){\makebox(0,0)[lb]{\smash{{\SetFigFont{12}{14.4}{\rmdefault}{\mddefault}{\updefault}{\color[rgb]{0,0,0}$\partial\Omega$}%
}}}}
\put(5101,-4111){\makebox(0,0)[lb]{\smash{{\SetFigFont{12}{14.4}{\rmdefault}{\mddefault}{\updefault}{\color[rgb]{0,0,0}$\Omega$}%
}}}}
\put(901,-5236){\makebox(0,0)[lb]{\smash{{\SetFigFont{12}{14.4}{\rmdefault}{\mddefault}{\updefault}{\color[rgb]{0,0,0}$\epsilon_0$}%
}}}}
\put(9151,-2011){\makebox(0,0)[lb]{\smash{{\SetFigFont{12}{14.4}{\rmdefault}{\mddefault}{\updefault}{\color[rgb]{0,0,0}$\epsilon_0$}%
}}}}
\put(3901,-1486){\makebox(0,0)[lb]{\smash{{\SetFigFont{12}{14.4}{\rmdefault}{\mddefault}{\updefault}{\color[rgb]{0,0,0}$C_0B^{-1/4}$}%
}}}}
\put(4426,-2461){\makebox(0,0)[lb]{\smash{{\SetFigFont{12}{14.4}{\rmdefault}{\mddefault}{\updefault}{\color[rgb]{0,0,0}$x_{j_{min}}=(0,0)$}%
}}}}
\put(2101,-3886){\makebox(0,0)[lb]{\smash{{\SetFigFont{12}{14.4}{\rmdefault}{\mddefault}{\updefault}{\color[rgb]{0,0,0}$x_j=(s_j,0)$}%
}}}}
\end{picture}%

}
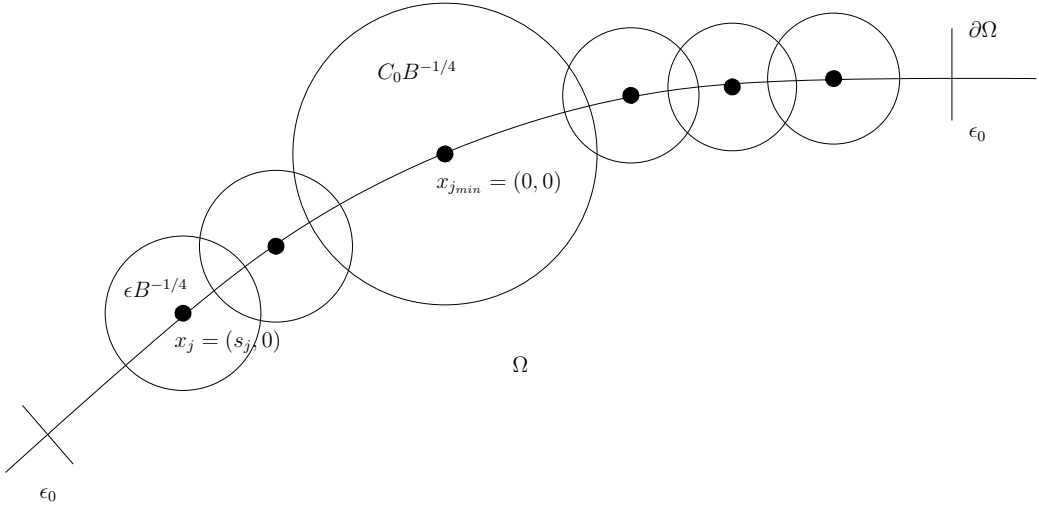
\caption{Partition of unity near the boundary}
\label{figurepartition}
\end{center}
\end{figure}
So, there exists $C>0$ such that for all $j$ :
$$\left|\int\Delta k_j(s)t (|\dr_t\tilde{v}_j|^2+|(i\dr_s+B\tilde{A}_1)\tilde{v}_j|^2) dsdt\right|\leq CB^{\frac{1}{4}}\|\tilde{v}_j\|^2.$$
We let :
\bg{equation}\label{alpha}
\alpha=\frac{1}{2}\frac{\dr^2\beta}{\dr s^2}(0,0).
\end{equation}
Using the assumption of non degeneracy of the minimum, we can choose $\eps_0>0$ small enough such that 
\bg{equation}\label{nondeg}
\frac{\alpha}{2}s^2\leq\tilde{\beta}(s,0)-\tilde{\beta}(0,0)\leq\frac{3}{2}\alpha s^2 
\end{equation}
for all $|s|\leq\eps_0$.\\
To estimate the other remainder, we will distinguish between three cases~:
\bg{list}{$\bullet$}{}
\item $j=j_{\min}$,
\item $|s_j|\geq\eps_0$,
\item $C_0B^{-1/4}\leq|s_j|\leq\eps_0$.
\end{list} 
\paragraph{Case 1 : $j=j_{min}$}~\\
As $$\frac{\dr\tilde{\beta}}{\dr s}(0,0)=0,$$
we have, with (\ref{approxbeta}) and (\ref{approxA}) :
$$|R_{j_{min}}(s,t)|\leq C (t^2+s^2t).$$ 
Consequently, using Proposition \ref{normalagmon}, we get :
$$\int|R_{j_{min}}(s,t)\tilde{v}_{j_{min}}|^2 dsdt\leq CB^{-2}\|\tilde{v}_{j_{min}}\|^2.$$
Taking $\la=B^{-1/2}$, we deduce :
$$q_{B\A}(v_{j_{min}}) \geq (\Theta_0 b'B-CB^{1/2})\|\tilde{v}_{j_{min}}\|^2.$$
\paragraph{Case 2 : $|s_j|\geq\eps_0$}~\\
We get :
$$|R_j(s,t)|\leq C((s-s_j)t+t^2).$$ 
Thus, we find :
$$\int|R_{j}(s,t)\tilde{v}_{j}|^2 dsdt\leq C (B^{-3/2}\eps^2+B^{-2})\|\tilde{v}_j\|^2.$$
Moreover, there exists $b''>b'$ such that for all $|s_j|\geq \eps_0$, we have : $\tilde{\beta_j}\geq b''$.
We take $\la=B^{-1/2}$ and deduce, using (\ref{diskcase}) and for $B$ large enough, that for all $j$ satisfying $|s_j|\geq\eps_0$ :
$$q_{B\A}(v_{j})\geq\Theta_0 b'B\|\tilde{v}_{j}\|^2.$$
\paragraph{Case 3 : $C_0B^{-1/4}\leq |s_j|\leq \eps_0$}~\\
We use the inequality :
$$\sup_{|s-s_j|\leq\eps B^{-1/4}}\left|\frac{\dr\tilde{\beta}}{\dr s}(0,s)\right|^2\leq C\sup_{|s-s_j|\leq\eps B^{-1/4}}|\tilde{\beta}-b'|$$
to find with (\ref{approxA}) and (\ref{approxbeta}) :
$$\int|R_{j}(s,t)\tilde{v}_{j}|^2 dsdt\leq C (B^{-3/2}\eps^2 \sup_{|s-s_j|\leq\eps B^{-1/4}}|\beta-b'|+B^{-2})\|\tilde{v}_j\|^2.$$
As a consequence, we can write, with $\la=B^{-1/2}$ :
$$q_{B\A}(v_{j})\geq (\Theta_0b'B+B(\Theta_0(\tilde{\beta}(s_j)-b')-C\eps^2\sup_{|s-s_j|\leq\eps B^{-1/4}}|\beta-b'|))\|\tilde{v}_j\|^2.$$
By non degeneracy, we have, for $C_0\geq 2\eps$ :
\bg{equation}\label{27}
\sup_{|s-s_j|\leq\eps B^{-1/4}}|\tilde{\beta}-b'|\leq 27 \inf_{|s-s_j|\leq\eps B^{-1/4}}|\tilde{\beta}-b'|.
\end{equation}
Indeed, we have, for all $C_0\geq 2\eps$ : 
$$\inf_{|s-s_j|\leq\eps B^{-1/4}}|\tilde{\beta}-b'|\geq\frac{\alpha}{2}\inf_{|s-s_j|\leq\eps B^{-1/4}} s^2\geq \frac{\alpha}{2}(s_j-\eps B^{-1/4})^2$$
and
$$\sup_{|s-s_j|\leq\eps B^{-1/4}}|\tilde{\beta}-b'|\leq\frac{3\alpha}{2}\sup_{|s-s_j|\leq\eps B^{-1/4}} s^2\leq \frac{3\alpha}{2}(s_j+\eps B^{-1/4})^2.$$
Thus, we get, for $C_0\geq 2\eps$ :
$$\ds{\frac{\sup_{|s-s_j|\leq\eps B^{-1/4}}|\tilde{\beta}-b'|}{\inf_{|s-s_j|\leq\eps B^{-1/4}}|\tilde{\beta}-b'|}\leq 3\left(\frac{s_j+\eps B^{-1/4}}{s_j-\eps B^{-1/4}}\right)^2}=3\left(1+\frac{2\eps B^{-1/4}}{s_j-\eps B^{-1/4}}\right)^2\leq 27.$$ 
We deduce, for $C_0\geq 2\eps$ :
$$q_{B\A}(v_{j})\geq (\Theta_0b'B+B(\Theta_0-27C\eps^2)\inf_{|s-s_j|\leq\eps B^{-1/4}}|\tilde{\beta}-b'|)\|\tilde{v_j}\|^2.$$
We will further use that there exists $c>0$ such that for all $C_0\geq 2\eps$ :
$$q_{B\A}(v_{j})\geq \left(\Theta_0b'B+cB(\tilde{\beta}(s_j)-b')\right)\|\tilde{v_j}\|^2.$$
Indeed, we have, for all $C_0\geq 2\eps$ : 
$$\inf_{|s-s_j|\leq\eps B^{-1/4}}|\tilde{\beta}-b'|\geq \frac{1}{27}(\tilde{\beta}(s_j)-b').$$
We find, for $\eps>0$ small enough :
$$q_{B\A}(v_{j})\geq (\Theta_0b'B+CB^{1/2})\|\tilde{v}_j\|^2.$$
We conclude that :
$$\sum_{j\,\,bnd} q_{B\A}(v_j)\geq (\Theta_0 b'B-CB^{1/2})\sum_{j\,\,bnd}\|v_j\|^2.$$
Putting together this estimate and the estimate inside $\Om$, we have the lower bound in Theorem \ref{roughestimate}.
\section{Models near a minimum of $\beta$ and upper bounds}
\subsection{Model operator}
We fix $k_0$, $k_1$ and $\alpha\geq 0$ and we wish to study the quadratic form on the Hilbert space $L^2((1-k_0 t)dtds)$ defined, for $u\in\mj{C}^{\infty}_0(B_{k_0})$ by :
\bg{equation}
q_{k_0,k_1,\alpha,B}(u)=\int_{\underset{0<t\leq\frac{1}{2k_0}}{s\in\R}}(1-tk_0)|\dr_t u|^2+(1-tk_0)^{-1}|(-i\dr_s+Bt(1-\frac{k_1}{2}t+\alpha s^2))u| dt ds,
\end{equation}
where $B_{k_0}=\R\times\left[0,\frac{1}{2k_0}\right[$ (and by convention $B_0=\R\times\R_+$).
The self-adjoint associated operator is :
$$-(1-k_0t)^{-1}\dr_t(1-k_0t)\dr_t+(1-tk_0)^2(-i\dr_s+Bt(1-\frac{k_1}{2}t+\alpha s^2))^2,$$
with Neumann condition on $t=0$ and Dirichlet condition on $t=\frac{1}{2k_0}$ (if $k_0\neq 0$). 
We first rescale the problem :
$$t=B^{-1/2}\tau,$$
$$s=B^{-1/4}\sigma,$$
and we are reduced to the operator on $L^2((1-tk_0B^{-1/2})dt ds)$ :
\bg{equation}\label{model1}
-\left(1-\frac{k_0t}{B^{1/2}}\right)^{-1}\dr_t\left(1-\frac{k_0t}{B^{1/2}}\right)\dr_t+\left(1-\frac{tk_0}{B^{1/2}}\right)^{-2}\left(t-\frac{k_1}{2B^{1/2}}t^2+\alpha \frac{s^2 t}{B^{1/2}}-i\frac{\dr_s}{B^{1/4}}\right)^2.
\end{equation}
We make a change of gauge $u\mapsto e^{i\xi_0 B^{1/4}\sigma}u$.
Then, the operator defined in (\ref{model1}) becomes :
\bg{equation}\label{model}
-\left(1-\frac{k_0t}{B^{1/2}}\right)^{-1}\dr_t\left(1-\frac{k_0t}{B^{1/2}}\right)\dr_t+\left(1-\frac{tk_0}{B^{1/2}}\right)^{-2}\left(t+\xi_0-\frac{k_1}{2B^{1/2}}t^2+\alpha \frac{s^2 t}{B^{1/2}}-i\frac{\dr_s}{B^{1/4}}\right)^2.
\end{equation}
\subsection{Degenerate case : $\alpha=0$}
This case corresponds to the degeneracy of the minimum of the restriction of $\beta$ to the boundary. In particular, we will prove Theorem \ref{constantfield}.
\subsubsection{Formal computation}
In order to have an upper bound, we first construct a formal quasimode.
We make a Fourier transform in the variable $s$. Thus, we are reduced to the study of the family of operators on $L^2((1-\frac{k_0t}{B^{1/2}})dt)$ :
$$
H_{k_0,k_1,\xi}=-(1-\frac{k_0t}{B^{1/2}})^{-1}\dr_t(1-\frac{k_0t}{B^{1/2}})\dr_t+(1-\frac{k_0t}{B^{1/2}})^{-2}(t+\xi-\frac{k_1}{2B^{1/2}}t^2)^2.
$$
We formally expand this operator in powers of $B$.\\
Term in $B^0$ : $$H_0=-\dr_t^2+(t+\xi)^2.$$
Term in $B^{-1/2}$ : $$H_1=k_0\dr_t-k_1(t+\xi)t^2+2k_0 t(t+\xi)^2.$$ 
We look for a quasimode expressed as :
$$\psi=\sum_{j=0}^{+\infty} B^{-j/2}u_j$$
and a expansion of the first eigenvalue :
$$\lambda_1(B)=\sum_{j=0}^{+\infty}\lambda_jB^{-j/2}.$$
So, we have to solve $$H_0u_0=\lambda_0 u_0$$
and, as we look for $\lambda_0$ minimal, we fix $\xi=\xi_0$, we deduce $\lambda_0=\Theta_0$ and we take $u_0=u_{\xi_0}$.
Then, the next equation to solve is :
$$H_0 u_1+H_1 u_0=\Theta_0 u_1+\lambda_1 u_0.$$ 
Thus, we deduce : $$(H_0-\Theta_0)u_1=(\lambda_1-H_1)u_0.$$
To have solutions, the second member must be orthogonal to $u_0$, so, using the formulas (\ref{formulasBS}), we get :
$$\lambda_1+\frac{k_0+k_1}{2}C_1-\Theta_0\xi_0(k_1-k_0)=0,$$
and we take : $$u_1=R_0 (\lambda_1-H_1)u_0.$$
We let :
$$\Theta_{1/2}^{k_0,k_1}=-\frac{k_0+k_1}{2}C_1+\Theta_0\xi_0(k_1-k_0).$$
Thus, $\psi$ is a good candidate to be a quasimode after truncation.
\subsubsection{Quasimode}
We write, in the initial coordinates (with $b'=1$, for simplicity) :
$$\tilde{A}_1=\conj{A}_1+R,$$
where $$\conj{A}_1=t(1-t\frac{k_1}{2})$$
with
\bg{equation}\label{k1} 
k_1=k_0-\frac{\dr\beta}{\dr t}(0,0).
\end{equation}
Let us denote $\psi=u_0+B^{-1/2}u_1$ and notice that $\psi$ is in the Schwartz class.
As a quasimode, we take :
$$u_B(s,t)=\chi(t)\psi(B^{1/2}t)e^{-s^2 B^{1/2-2\rho}}e^{i\xi_0 B^{1/2}s},$$
with $\chi$ a smooth cutoff function supported in $\left[0,\frac{1}{2k_0}\right]$ and $\rho\in]0,\frac{1}{4}[$ which will be choosen later to optimize the error. The Gaussian $e^{-s^2B^{1/2-2\rho}}$ permits a localization near $s=0$.
We have :
\bg{eqnarray*}
q_{B\A}(u_B)\leq \int (1-tk_0)|\dr_t u_B|^2+(1-tk_0)^{-1}|(-i\dr_s+B\tilde{A}_1)u_B|^2 ds dt\\
+C\int\Delta k(s)t\{|\dr_t u_B|^2+|(-i\dr_s+B\tilde{A}_1)u_B|^2\}ds dt.
\end{eqnarray*}
By noticing that there exists $C>0$ such that :
$$|\tilde{A}_1(s,t)|\leq C t,$$
we get :
$$\left|\int\Delta k(s)t\{|\dr_t u_B|^2+|(-i\dr_s+B\tilde{A}_1)u_B|^2\}ds dt\right|\leq CB^{1/4+\rho}\|u_B\|^2.$$
Let us prove the upper bound for the first term (the second can be treated in the same way). We have :
$$\dr_t u_B=\chi'(t)\psi(B^{1/2}t)e^{-s^2 B^{1/2-2\rho}}e^{i\xi_0 B^{1/2}s}+B^{1/2}\chi(t)\psi'(B^{1/2}t)e^{-s^2 B^{1/2-2\rho}}e^{i\xi_0 B^{1/2}s}.$$
Thus we get :
$$|\dr_t u_B|^2\leq 2|\chi'(t)\psi(B^{1/2}t)|^2e^{-2s^2 B^{1/2-2\rho}}+2B|\chi(t)\psi'(B^{1/2}t)|^2e^{-2s^2 B^{1/2-2\rho}}.$$
Then, we find : 
\bg{align*}
\int t\Delta k(s)|\dr_t u_B|^2 dt ds&\leq& C\int ts|\chi'(t)|^2|\psi(B^{1/2}t)|^2e^{-2s^2 B^{1/2-2\rho}} dt ds\\
                                    &  + & CB\int ts |\chi(t)|^2|\psi'(B^{1/2}t)|^2e^{-2s^2 B^{1/2-2\rho}} dt ds. 
\end{align*}
As $\psi$ is in the Schwartz class, we get :
$$\int ts|\chi'(t)|^2|\psi(B^{1/2}t)|^2e^{-2s^2 B^{1/2-2\rho}} dt ds=O(B^{-\infty})\|u_B\|^2.$$
Then, we have after rescaling, for some $C>0$ independent of $B$ :
$$B\int ts |\chi(t)|^2|\psi'(B^{1/2}t)|^2e^{-2s^2 B^{1/2-2\rho}} dt ds\leq C B B^{-1/2} B^{-1/4+\rho}\|u_B\|^2=CB^{1/4+\rho}\|u_B\|^2.$$
Moreover, we have :
\bg{eqnarray*}
\int (1-tk_0)|\dr_t u_B|^2+(1-tk_0)^{-1}|(-i\dr_s+B\tilde{A}_1)u_B|^2 ds dt\\
=\int (1-tk_0)|\dr_t u_B|^2+(1-tk_0)^{-1}|(-i\dr_s+B\conj{A}_1)u_B|^2 ds dt\\
+\Re\left\{\int (1-tk_0)^{-1}(B^2|Ru_B|^2+2B(-i\dr_s+B\conj{A}_1)u_B Ru_B)ds dt\right\}.
\end{eqnarray*}
We get :
$$q_{k_0,k_1,0,B}(u_B)\leq (\Theta_0 B+\Theta_{1/2}^{k_0,k_1}B^{1/2}+CB^{1/2-2\rho})\|u_B\|^2,$$
the crucial points being to estimate the term 
$$\int |\dr_s^2 e^{-s^2B^{1/2-2\rho}}|e^{-s^2B^{1/2-2\rho}}|\chi(t)|^2|\psi(B^{1/2}t)|^2 dt ds$$ by $O(B^{1/2-2\rho})\|u_B\|^2$ and the term $\int (Bt+B^{1/2}\xi_0)\dr_s(\chi(t)\psi(B^{1/2}t)e^{-s^2B^{1/2-2\rho}})$ by $O(B^{-\infty})\|u_B\|^2$ thanks to the fact that $M_1=0$ (cf. (\ref{formulasBS})) and that $\psi$ is in the Schwartz class. 
Using that :
$$|R(s,t)|\leq C(t^3+s^4 t+st^2),$$
we find :
$$\left|\Re\left\{\int (1-tk_0)^{-1}(B^2|Ru_B|^2+2B(-i\dr_s+B\conj{A}_1)u_B Ru_B)ds dt\right\}\right|\leq CB^{1/4+\rho}\|u_B\|^2.$$
and finally with $\rho=\frac{1}{12}$ :
$$q_{B\A}(u_B)\leq (\Theta_0 B+\Theta_{1/2}^{k_0,k_1}B^{1/2}+CB^{1/3})\|u_B\|^2.$$
Thus, after replacing $k_1$ by its expression, the upper bound of Theorem \ref{constantfield} is proved.\\
\bg{rem}
It follows from the identities (\ref{formulasBS}) that : 
$$\frac{C_1}{2}-\Theta_0\xi_0=M_3-\xi_0^3>0,$$
where $\ds{M_3=\int_{t>0}(t+\xi_0)^3 u_0^2 dt.}$
This remark permits to understand how the upper bound of Theorem \ref{constantfield} improves the one of Aramaki. 
\end{rem} 
\subsection{Non-degenerate case $\alpha>0$}
\subsubsection{Formal computation}\label{formal}
We consider the operator $H$ (cf.(\ref{model})) :
$$
-(1-\frac{k_0t}{B^{1/2}})^{-1}\dr_t(1-\frac{k_0t}{B^{1/2}})\dr_t+(1-\frac{k_0t}{B^{1/2}})^{-2}(t+\xi_0-\frac{k_1}{2B^{1/2}}t^2+\frac{\alpha}{B^{1/2}} s^2 t-i\frac{\dr_s}{B^{1/4}})^2.
$$
Formally, we write : 
$$H=\sum_{j=1}^{+\infty}B^{-j/4}H_j.$$
Let us look for a quasimode expressed as :
\bg{equation}\label{quasimode}
U=\sum_{j=1}^{+\infty}B^{-j/4}U_j.
\end{equation}
and a Taylor expansion of the lowest eigenvalue :
$$\lambda_1^N(B)=\sum_{j=1}^{+\infty}\Theta_{j/4} B^{-j/4}.$$
Here, we have :
$$H_0=-\dr_t^2+(t+\xi_0)^2,$$
$$H_1=-2i\dr_s(t+\xi_0),$$
$$H_2=k_0\dr_t-\dr_s^2+2(t+\xi_0)(\alpha s^2 t-\frac{k_1}{2}t^2)+2k_0 t(t+\xi_0)^2.$$
This leads us to solve :
$$H_0 U_0=\lambda_0 U_0.$$
We write $U_0$ as $U_0=u_0(t)\psi_0(s)$ and, as we look for $\lambda_1^N$ minimal, we take $\lambda_0=\Theta_0$ and $u_0>0$ the associated normalized eigenvector.\\
Then, we solve : 
$$H_1 U_0+H_0 U_1=\Theta_0 U_1+\lambda_1 U_0.$$
We can take $\Theta_{1/4}=0$ by writing $U_1=u_1(t)\psi_1(s)$ with $\psi_1=\dr_s\psi_0$ and we find :
$$(H_0-\Theta_0)u_1=2i(t+\xi_0)u_0.$$
As $M_1=0$ (see (\ref{formulasBS})), this last equation admits a unique solution $u_1$ such that $\int_{t>0}u_0 u_1 dt=0$.\\
Finally, we consider :
$$H_0 U_2+H_1 U_1+H_2 U_0=\Theta_0 U_2+\Theta_{1/2} U_0.$$
Thus, we get :
$$(H_0-\Theta_0)U_2=-H_1 U_1-H_2 U_0+\Theta_{1/2} U_0=2i(t+\xi_0)u_1\dr_s\psi_1-H_2 U_0+\Theta_{1/2} U_0.$$
Multiplying by $u_0$ and integrating with respect to $t$, one applies the formulas (\ref{formulasBS}) and one solves :
$$-(1-4I_2)\dr_s^2\psi_0+\alpha\Theta_0 s^2 \psi_0=\left(\Theta_{1/2}+\frac{k_0+k_1}{2}C_1-(k_1-k_0)\Theta_0\xi_0\right)\psi_0.$$
where
$$I_2=\int_{t>0}(t+\xi_0)R_0((t+\xi_0)u_0)u_0 dt.$$
This last integral can be rewritten by letting $v=R_0((t+\xi_0)u_0)$ ; we have : $$(H_0-\Theta_0)v=(t+\xi_0)u_0.$$
By computing, we get :
$$-\frac{1}{2}\frac{\dr u}{\dr\xi}(\cdot,\xi_0)=v.$$
Using the identities of \cite{FouHel2}, we find :
$$1-4I_2=\frac{\mu''(\xi_0)}{2}=3C_1\sqrt{\Theta_0}>0.$$
After rescaling, we let :
$$\ds{\psi_0(s)=e^{-\frac{\Theta_0^{1/4}\sqrt{\alpha}s^2}{2\sqrt{3C_1}}}}$$
and :
\bg{equation}\label{theta12}
\Theta_{1/2}=\Theta_{1/2}^{k_0,k_1,\alpha}=-\frac{k_0+k_1}{2}C_1+(k_1-k_0)\Theta_0\xi_0+\sqrt{3C_1}\Theta_0^{3/4}\sqrt{\alpha}.
\end{equation}
\subsubsection{Quasimode}\label{approximation}
For simplicity, we assume $b'=1$.
We write : 
$$\tilde{A}_1=\conj{A}_1+R,$$ where 
$$\conj{A}_1=t(1-t\frac{k_1}{2}+\alpha s^2)$$
with $\alpha$ defined in (\ref{alpha}) and $k_1$ defined in (\ref{k1}).
We let : 
$$u_B(s,t)=\chi(t)U(B^{1/4}s,B^{1/2}t)e^{i\xi_0 B^{1/2}s},$$
where $U$ consists of the three first terms of (\ref{quasimode}). 
We have :
\bg{eqnarray*}
q_{B\A}(u_B)\leq \int (1-tk_0)|\dr_t u_B|^2+(1-tk_0)^{-1}|(-i\dr_s+B\tilde{A}_1)u_B|^2 ds dt\\
+C\int\Delta k(s)t\{|\dr_t u_B|^2+|(-i\dr_s+B\tilde{A}_1)u_B|^2\}ds dt.
\end{eqnarray*}
Moreover, we have :
\bg{eqnarray*}
\int (1-tk_0)|\dr_t u_B|^2+(1-tk_0)^{-1}|(-i\dr_s+B\tilde{A}_1)u_B|^2 ds dt\\
=\int (1-tk_0)|\dr_t u_B|^2+(1-tk_0)^{-1}|(-i\dr_s+B\conj{A}_1)u_B|^2 ds dt\\
+\Re\left\{\int (1-tk_0)^{-1}(B^2|Ru_B|^2+2B(-i\dr_s+B\conj{A}_1)u_B Ru_B)ds dt\right\}.
\end{eqnarray*}
Using that $U$ is in the Schwartz class, we get :
$$q_{k_0,k_1,\alpha,B}(u_B)\leq (\Theta_0 B+\Theta_{1/2}^{k_0,k_1,\alpha}B^{1/2}+C)\|u_B\|^2.$$
Moreover, we have :
$$|\tilde{A}_1-\conj{A}_1|\leq C(s^3t+st^2+t^3).$$
So, we get :
$$\left|\Re\left\{\int (1-tk_0)^{-1}(B^2|Ru_B|^2+2B(-i\dr_s+B\conj{A}_1)u_B Ru_B)ds dt\right\}\right|\leq CB^{1/4}\|u_B\|^2.$$
Finally, we find :
$$q_{B\A}(u_B)\leq (\Theta_0 B+\Theta_{1/2}^{k_0,k_1,\alpha}B^{1/2}+CB^{1/4})\|u_B\|^2.$$
In particular, we have proved the upper bound in Theorem \ref{roughestimate}.

\section{Tangential Agmon's estimates}
We first observe that, for $\Phi$ a real Lipschitzian function and if $u$ is in the domain of the Neumann realization of $(i\nabla+B\A)^2$, then we have, by integration by parts :
$$\Re\langle(i\nabla+B\A)^2u,\exp(2B^{1/2}\Phi)u\rangle=q_{B\A}(\exp(B^{1/2}\Phi)u)-B\||\nabla\Phi|\exp(B^{1/2}\Phi)u\|^2.$$
Taking $u=u_B$ an eigenfunction attached to the lowest eigenvalue $\la^1(B\A)$~, we get :
\bg{equation}\label{agmon}
\la^1(B\A)\|\exp(B^{1/2}\Phi)u_B\|^2=q_{B\A}(\exp(B^{1/2}\Phi)u_B)-B\||\nabla\Phi|\exp(B^{1/2}\Phi)u_B\|^2.
\end{equation}
\subsection{Tangential Agmon's estimates for $u_B$}
We now use the lower bound found in Section 2 ; more precisely, for all $\eps>0$, there exists $c>0$ and $C>0$ such that, for all $C_0>0$ sufficiently large, there exists $C'>0$ s.t for all $u$ in the form domain of $q_{B\A}$ :
\bg{eqnarray*}
q_{B\A}(u)&\geq& (bB-CB^{1/2})\sum_{j\,\,int}\|\chi_j u\|^2\\
&&+\sum_{j\,\,bnd, j\neq j_{min}}(\Theta_0 b'B+c(\beta(s_j)-b')B)\|\chi_j u\|^2\\
&&+(\Theta_0 b'B-C'B^{1/2})\|\chi_{j_{min}}u\|^2.
\end{eqnarray*}
We choose $u=\exp(B^{1/2}\Phi)u_B$ ; we recall that, by Theorem \ref{roughestimate}, we have the upper bound :
$$\la^1(B\A)\leq \Theta_0 b'B+CB^{1/2}.$$
Using these estimates in (\ref{agmon}), we find the inequality by dividing by $B$ :
\bg{eqnarray*}
\int (C'B^{-1/2}+|\nabla\Phi|^2)|\chi_{j_{min}}\exp(B^{1/2}\Phi)u_B|^2\geq\\ 
\sum_{\stackrel{j\,\,bnd}{j\neq j_{min}}}\int (c(\tilde{\beta}(s_j)-b')-CB^{-1/2}-|\nabla\Phi|^2)|\chi_j \exp(B^{1/2}\Phi)u_B|^2 ds dt.
\end{eqnarray*}
We choose 
$$\Phi=\alpha_1 d(s),$$ 
where $d$ is the Agmon distance associated with the metric $(\beta(s,0)-b') ds^2$ i.e :
$$d(s)=\int_{0}^{|s|}(\beta(\sigma,0)-b')^{1/2}d\sigma.$$
On $D_{j_{min}}$, we notice that 
$$|\nabla\Phi|^2\leq C B^{-1/2}.$$
Then, for $j\neq j_{min}$, we consider the quantity :
$$c(\tilde{\beta}(s_j)-b')-CB^{-1/2}-\alpha_1^2(\tilde{\beta}(s)-b').$$
For $\eps>0$ and $\alpha_1$ small enough, there exists $c'>0$ such that for $j$ such that $|s_j|\geq\eps_0$ and $B$ large enough, we have :
$$c(\tilde{\beta}(s_j)-b')-CB^{-1/2}-\alpha_1^2(\tilde{\beta}(s)-b')\geq c'.$$
For $C_0\geq 2\eps$, there exists $c''>0$ such that for $j\neq j_{min}$ and $|s_j|\leq\eps_0$and $B$ large enough, we have :
$$c(\tilde{\beta}(s_j)-b')-CB^{-1/2}-\alpha_1^2(\tilde{\beta}(s)-b')\geq c''B^{-1/2}.$$
Indeed, due to the non degeneracy, we have (\ref{27}).
Thus, we get $C>0$ and $B_0>0$ such that for all $B\geq B_0$ :
$$\sum_{j\,\,bnd}\int |\chi_j \exp(B^{1/2}\Phi)u_B|^2\leq C\int_{|s|\leq C_0 B^{-1/4}}|\exp(B^{1/2}\Phi)u_B|^2 .$$
We deduce Proposition \ref{tagm1} and have the following corollary :
\bg{cor}
For all $n\in\N$, there exists $C>0$ such that for all $B$ large enough :
$$\int_{\Om}s^{2n}\{|u_B|^2+B^{-1}|(i\nabla+B\A)u_B|^2\} dx \leq CB^{-n/2}\int_{\Om}|u_B|^2 dx.$$
\end{cor}
\subsection{Agmon's estimates for $D_s u_B$}
We consider a partition of unity as in (\ref{partition}).
We have the formula (\ref{IMS}) and :
$$q_{B\A}(u)\geq \sum_{j} q_{B\A}(\chi_j^B u)-CB^{1/2}\|u\|^2.$$
We use (\ref{agmon}).
We have $$\la^1(B)\leq \Theta_0 b'B+CB^{1/2}.$$
Thus, we get, using the inequalities of the previous section :
\bg{eqnarray*}
q_{B\A}\left(\chi_{j_{min}}e^{B^{1/2}\Phi}u_B\right)+\sum_{j\neq j_{min}}(\Theta_0 b'B+c(\beta(s_j)-b')B)\left|\chi_j e^{B^{1/2}\Phi}u_B\right|^2\\
+bB\sum_{j\,\,int}\left|\chi_j e^{B^{1/2}\Phi}u_B\right|^2\\
\leq (\Theta_0b'B+CB^{1/2})\int\left|e^{B^{1/2}\Phi} u_B\right|^2+B\int\left|\nabla\Phi e^{B^{1/2}\Phi} u_B\right|^2+CB^{1/2}\|u_B\|^2,
\end{eqnarray*}
where $\Phi=\alpha_1 d(s)$.\\
We have the control :
$$B\int_{\Om}\left|\chi_{j_{min}}\nabla\Phi e^{B^{1/2}\Phi} u_B\right|^2\leq CB^{1/2}\int_{\Om}\left|\chi_{j_{min}}e^{B^{1/2}\Phi} u_B\right|^2\leq CB^{1/2}\int_{\Om}|u_B|^2 dx,$$
and we deduce, for $\alpha_1$ small enough :
$$q_{B\A}(\chi_{j_{min}}e^{B^{1/2}\Phi}u_B)-\Theta_0b'B \left|\chi_{j_{min}}e^{B^{1/2}\Phi}u_B\right|^2\leq C B^{1/2}\int |u_B|^2.$$
We introduce :
\bg{align}\label{qapp}
&q_{app}(v)=\int_{t>0,s\in\R}(1-k_0 t)|\dr_t v|^2+\\
&(1-k_0 t)^{-1}|(Bt+B\alpha s^2 t+B^{1/2}\xi_0-D_s-B\frac{k_1}{2}t^2)v|^2dt ds.\nonumber
\end{align}
If we write :
$$\tilde{A}_1(s,t)=\int (1-t'k(s))\tilde{\beta}(s,t') dt',$$
we have :
$$(1-tk(s))\tilde{\beta}(s,t)=(1-tk_1)+\alpha s^2+O(t^2+st+s^3),$$
and thus :
\bg{equation}\label{A1tilde}
\tilde{A}_1(s,t)=t-\frac{k_1}{2}t^2+\alpha s^2 t+O(t^3+st^2+s^3 t).
\end{equation}
Then, by the Cauchy-Schwarz inequality, we have for all $\la>0$ :
$$q_{B\A}(v)\geq (1-\la)q_{app}(v)-\frac{B^2}{\la}\|Rv\|^2.$$
For instance, we can estimate $\int (st^2)^2 |v|^2$. 
Using the tangential (cf. Proposition \ref{tagm1}) and normal Agmon estimates and letting : 
$$v=\chi_{j_{min}}e^{B^{1/2}\Phi}u_B,$$
we have :
$$B^2\int s^2 t^4 |v|^2 dsdt\leq CB^2B^{-1/2}B^{-2}\|v\|^2.$$ 
In the same way, we control the other remainders and by choosing $\la$ correctly, we get :
\bg{equation}\label{remainder}
q_{B\A}(v)\geq q_{app}(v)-CB^{1/4}\int |v|^2.
\end{equation}
Using the Cauchy-Schwarz inequality and again the Agmon estimates, we find :
$$q_{app}(v)\geq (1-B^{-1/2})q_{app}^2(v)-CB^{1/2}\int|v|^2,$$
where 
$$q_{app}^2(v)=\int_{t>0,s\in\R}(1-k_0 t)|\dr_t v|^2+(1-k_0 t)^{-1}|(Bt+B^{1/2}\xi_0-D_s)v|^2 dx.$$
Making a Fourier transform in the variable $s$ and letting $w=\hat{v}$, we have :
$$q_{app}^2(v)=\int_{t>0,\sigma\in\R}(1-k_0 t)|\dr_t w|^2+(1-k_0 t)^{-1}|(Bt+B^{1/2}\xi_0-\sigma)w|^2 dt d\sigma.$$
Thus, we get (see \cite[Chapter 6, Prop 6.2.1]{FouHel2} or \cite[Section 11]{HelMo3}) :
$$q_{app}^2(v)\geq \Theta_0b'B\int|v|^2+B^{1/2}\frac{\mu''(\xi_0)}{2} \int|D_s v|^2-CB^{1/2}\int|v|^2.$$
Consequenlty, we get the upper bound :
$$\int \left|D_s(\chi_{j_{min}}e^{B^{1/2}\Phi}u_B)\right|^2\leq C\int|u_B|^2.$$
We deduce the following proposition :
\bg{prop}[Tangential Agmon's estimates for $D_s u_B$]\label{tagm2}
With the previous notations, there exists $C>0$ and $\alpha_1>0$ such that for all $B$ large enough : 
$$\int_{\Om}\left|e^{\alpha_1 B^{1/2}\chi(t(x))d(s(x))}D_s u_B\right|^2 dx\leq CB^{1/2}\int_{\Om}|u_B|^2 dx,$$
where $\chi$ is a smooth cutoff function supported in $[-t_0,t_0]$.
\end{prop}
\bg{cor}
For all $n\in\N$, there exists $C>0$ such that for all $B$ large enough, we have :
$$\int_{\Om} \chi(t)s^{2n} |D_s u_B|^2 dx\leq CB^{1/2-n/2}\int_{\Om}|u_B|^2 dx.$$
\end{cor}
\bg{rem}
The tangential and normal Agmon estimates roughly say that $|u_B|$ has the same behaviour as $e^{-\alpha s^2B^{1/2}}u_0(B^{1/2}t)$.
\end{rem}
\section{Refined lower bounds}
In this section, we prove the lower bound in Theorem \ref{preciseestimate}.
We consider a partition of unity as in (\ref{partition}) with $\ds{\rho=\frac{1}{4}-\eta}$ for $\eta>0$. 
We have :
$$q_{B\A}(u)\geq \sum_{j} q_{B\A}(\chi_j^B u)-CB^{1/2-2\eta}\|u\|^2.$$
\subsection{Control far from the minimum}
Let us first recall some the estimates we have proved.
For $j$ such that $D_j$ does not intersect the boundary, we have :
$$q_{B\A}(\chi_j u)\geq bB\int |\chi_j u|^2 dx.$$
For $j$ such that $D_j$ intersect the boundary and $j\neq j_{min}$, we notice that, for $B$ large enough :
$$q_{B\A}(\chi_j u)\geq \Theta_0 b'B\int|\chi_j u|^2.$$
\subsection{Reduction to a model near the minimum}
Using the inequalities of the previous section, we get :
$$q_{B\A}(u_B)\geq \Theta_0 b' B\sum_{j\neq j_{min}}\|\chi_j u_B\|^2+q_{B\A}(\chi_{j_{min}}u_B)-CB^{1/2-2\eta}\|u_B\|^2.$$
By the normal and tangential Agmon estimates, we have proved in (\ref{remainder}), with (\ref{qapp}), (\ref{A1tilde}) and the Cauchy-Schwarz inequality :
$$q_{B\A}(\chi_{j_{min}}u_B)\geq q_{app}(\chi_{j_{min}}u_B)-CB^{1/4}\|u_B\|^2.$$
In order to make the term in $\alpha s^2 t$ disappear, we make the change of variables~: 
$$t=\la(s)\tau,$$
where $\la(s)=(1+\alpha s^2)^{-1/2}$ ; we have 
\bg{eqnarray}\label{newderivatives}
\dr_s v=\frac{\dr\tau}{\dr s}\dr_{\tau}\tilde{v}+\dr_s\tilde{v},&\ds{\dr_t v=\frac{\dr\tau}{\dr t}\dr_{\tau}\tilde{v}},
\end{eqnarray}
where $\tilde{v}$ denotes the function $v$ in the variables $(\tau,s)$ and we are reduced to the form :
\bg{eqnarray*}
\widetilde{q_{app}}(v)&=&\int \Big\{\big(1-k_0 \tau\la(s)\big)|\dr_\tau v|^2\\
&+&\big(1-k_0 \tau\la(s)\big)^{-1}|(B\tau+\xi_0\la(s)B^{1/2}-\la(s)D_s-B\frac{k_1 \tau^2}{2}\la(s)^3\\
&+&\alpha\tau s\la(s)^3D_{\tau})v|^2\Big\}\la(s)^{-1}d\tau ds,
\end{eqnarray*}
where we have omitted the tilde.
Noticing that $s^2=O(B^{2\rho-1/2})$, on the support of $v=\chi_{j_{min}} u_B$, we make the approximations in $L^2$ : 
$$-\la(s)D_s v=-D_s v+O\left(s^2\right)D_s v,$$
$$\tau^2\la(s)^3v=\tau^2v+O\left(s^2\tau^2\right)v,$$
$$s\la(s)^3\tau D_\tau v=s\tau D_{\tau}v+O(s^3\tau) D_{\tau}v.$$
We first find :
\bg{align}
&\widetilde{q_{app}}(v)\geq\int\big\{(1-\tau k_0)|\dr_\tau v|^2\nonumber\\
&+(1-\tau k_0)^{-1}\big|\big(B\tau+\xi_0\la(s)B^{1/2}-\la(s)D_s-B\frac{k_1 \tau^2}{2}\la(s)^3+\alpha s\la(s)^3\tau D_\tau\big)v\big|^2\big\}\la(s)^{-1}d\tau ds\nonumber\\
&-C\int\Delta\la(s)\tau\big\{|\dr_\tau v|^2+\big|\big(B\tau+\xi_0\la(s)B^{1/2}-\la(s)D_s-B\frac{k_1 \tau^2}{2}\la(s)^3\nonumber\\
&+\alpha s\la(s)^3\tau D_\tau\big)v\big|^2\big\}\la(s)^{-1}d\tau ds,\nonumber
\end{align}
where $$\Delta\la(s)=\la(s)-\la(0).$$
Let us consider the second term :
\bg{align*}
&\int\Delta\la(s)\tau\big\{|\dr_\tau v|^2+\big|\big(B\tau+\xi_0\la(s)B^{1/2}-\la(s)D_s-B\frac{k_1 \tau^2}{2}\la(s)^3\\
&+\alpha s\la(s)^3\tau D_\tau\big)v\big|^2\big\}\la(s)^{-1}d\tau ds.
\end{align*}
Coming back in the variables $(t,s)$, this term becomes :
\bg{align*}
\int\frac{\Delta\la(s)}{\la(s)} t\big\{|\dr_t v|^2+|(B(1+\alpha s^2)t+\xi_0 B^{1/2}-D_s-B\frac{k_1 t^2}{2})v|^2\big\}dt ds.
\end{align*}
Thus, the Agmon estimates give a control of the second term of order $O(1)$.
Then, by the Cauchy-Schwarz inequality, the Agmon estimates (for $u_B$ and $D_s u_B$ after having come back in the variables $(s,t)$) and using the same kind of analysis as in (\ref{remainder}), we have :
\bg{align}
&\int\big\{(1-\tau k_0)|\dr_{\tau} v|^2+(1-\tau k_0)^{-1}\big|\big(B\tau+\xi_0\la(s)B^{1/2}-\la(s)D_s-B\frac{k_1 \tau^2}{2}\la(s)^3\nonumber\\
&+\alpha s\la(s)^3\tau D_\tau\big)v\big|^2\big\}\la(s)^{-1}d\tau ds\nonumber\\
&\geq \int\big\{(1-\tau k_0)|\dr_\tau v|^2+(1-\tau k_0)^{-1}\big|\big(B\tau+\xi_0\la(s)B^{1/2}-D_s-B\frac{k_1 \tau^2}{2}\big)v\big|^2\big\}\la(s)^{-1}d\tau ds\nonumber\\
&-CB^{1/4}\|v\|^2.\nonumber
\end{align}
We have finally, with $v=\chi_{j_{min}}u_B$ :
\bg{align}\label{redminimum}
&q_{B\A}(u_B)\geq \Theta_0 b'B\sum_{j\neq j_{min}}\|\chi_j u_B\|^2\\
&+\int\big\{(1-\tau k_0)|\dr_\tau v|^2+(1-\tau k_0)^{-1}\big|\big(B\tau+\xi_0\la(s)B^{1/2}-D_s-B\frac{k_1 \tau^2}{2}\big)v\big|^2\big\}\la(s)^{-1}d\tau ds\nonumber\\
&-CB^{1/4}\|u_B\|^2-CB^{1/2-2\eta}\|u_B\|^2.\nonumber
\end{align}
Moreover, thanks to the exponential decrease of $u_B$ away from the boundary (normal Agmon estimates), we can replace $\chi_{j_{min}}$ by a smooth cutoff function such that $$\su \chi_{j_{min}}\subset \{0<t\leq B^{-1/2+\eta}\,\mathrm{and}\,|s|\leq B^{-1/4+\eta}\},$$
that is we assume $\chi_{j_{min}}$ is supported in rectangles rather than balls ; the reason is technical and will appear in the next section. 
\subsection{Lower bound for the model}
So, we are reduced, after the rescaling $\ds{\tau=\frac{\hat{\tau}}{B^{1/2}}}$, $\ds{s=\frac{\hat{s}}{B^{1/4}}}$,
to the study of : 
\bg{align*}
&q_{mod}(u)=\int_{\hat{\tau}>0,\hat{s}\in\R}\Big\{(1-\frac{k_0 \hat{\tau}}{B^{1/2}})|\dr_{\hat{\tau}} u|^2\\
&+(1-\frac{k_0 \hat{\tau}}{B^{1/2}})^{-1}|(\hat{\tau}+\xi_0\la(B^{-1/4}\hat{s})-\frac{D_{\hat{s}}}{B^{1/4}}-\frac{k_1}{2 B^{1/2}}\hat{\tau}^2)u|^2\Big\}(1+\frac{\alpha \hat{s}^2}{B^{1/2}})^{1/2}d\hat{\tau} d\hat{s}.
\end{align*}
\paragraph{Reduction to the euclidean measure}~\\
In order to make disappear the measure $(1+\frac{\alpha \hat{s}^2}{B^{1/2}})^{1/2}$, we make the change of function defined by :
$$v=\left(1+\frac{\alpha \hat{s}^2}{B^{1/2}}\right)^{1/4}u=f_B(\hat{s})u,$$
we have :
\bg{align*}
&q_{mod}(u)=\int_{\hat{\tau}>0,\hat{s}\in\R}\Big\{(1-\frac{k_0 \hat{\tau}}{B^{1/2}})|\dr_{\hat{\tau}} v|^2\\
&+(1-\frac{k_0 \hat{\tau}}{B^{1/2}})^{-1}|(\hat{\tau}+\xi_0\la(B^{-1/4}\hat{s})-\frac{D_{\hat{s}}}{B^{1/4}}-\frac{f'_B(\hat{s})}{B^{1/4}f_B(\hat{s})}-\frac{k_1}{2 B^{1/2}}\hat{\tau}^2)v|^2\Big\}d\hat{\tau} d\hat{s}.
\end{align*}
\paragraph{Term in $\hat{s}$}~\\
We want to make a Fourier transform in the variable $\hat{s}$ to be reduced to a problem on a half axis, but the term
$\xi_0\la(B^{-1/4}\hat{s})$ is annoying ; that is why we make it disappear with a change of gauge.
We write : $\la(B^{-1/4}\hat{s})=1+r_B(\hat{s})$ and we make the change of gauge
${v\mapsto\tilde{v}=ve^{-i\phi(\hat{s})}}$, where 
$$\phi(\hat{s})=\int_0^{\hat{s}} \xi_0 r_B(\sigma)-\frac{1}{B^{1/4}}\frac{f'_B(\sigma)}{f_B(\sigma)}d\sigma$$  
to be reduced to :
\bg{align*}
&\widetilde{q_{mod}}(\tilde{v})=\int_{\hat{\tau}>0,\hat{s}\in\R}\Big\{(1-\frac{k_0 \hat{\tau}}{B^{1/2}})|\dr_{\hat{\tau}} \tilde{v}|^2\\
&+(1-\frac{k_0 \hat{\tau}}{B^{1/2}})^{-1}|(\hat{\tau}+\xi_0-\frac{D_{\hat{s}}}{B^{1/4}}-\frac{k_1}{2 B^{1/2}}\hat{\tau}^2)\tilde{v}|^2\Big\}d\hat{\tau} d\hat{s},
\end{align*}
where $u=(\chi_{j_{min}}u_B)(B^{1/2}\hat{\tau},B^{1/4}\hat{s}).$ 
We make a Fourier transform in the variable $\hat{s}$ and we are reduced to a half axis problem in the normal variable~:
$$q_n(w)=\int_{\hat{\tau}>0}(1-\frac{k_0 \hat{\tau}}{B^{1/2}})|\dr_{\hat{\tau}} w|^2+(1-\frac{k_0 \hat{\tau}}{B^{1/2}})^{-1}|(\hat{\tau}+\xi_0-\frac{\sigma}{B^{1/4}}-\frac{k_1}{2 B^{1/2}}\hat{\tau}^2)w|^2 d\hat{\tau},$$
with $w=\hat{v}$.
\paragraph{Model on a half axis}~\\
We can apply the same kind of analysis as in \cite[Chapter 6, Prop 6.2.1]{FouHel2} or in \cite[Section 11]{HelMo3} to get the lower bound ; there exists $C>0$ such that for all $B$ large enough :
\bg{equation}\label{bottommodel}
q_n(w)\geq\left(\Theta_0+(\Theta_{1/2}^{k_0,k_1}+\frac{\mu''(\xi_0)}{2}\sigma^2)B^{-1/2}-CB^{-3/4+3\eta}\right)\int_{\hat{\tau}>0}|w|^2\left(1-\frac{\hat{\tau}k_0}{B^{1/2}}\right)d\hat{\tau}.
\end{equation}
\bg{rem}
In \cite{FouHel2}, the fact that the magnetic field is constant permits to be reduced to the case $k_0=k_1=1$, thus $\ds{\Theta_{1/2}^{k_0,k_1}=-C_1}$. 
\end{rem}
Let us just recall the main ideas of the proof.
We consider first the (formal) operator on $L^2((1-\frac{k_0\hat{\tau}}{B^{1/2}})d\hat{\tau})$ :
$$\got{h}(\sigma,B)=-(1-\frac{k_0\hat{\tau}}{B^{1/2}})^{-1}\frac{d}{d\hat{\tau}}(1-\frac{k_0\hat{\tau}}{B^{1/2}})\frac{d}{d\hat{\tau}}+(1-\frac{k_0\hat{\tau}}{B^{1/2}})^{-2}(\hat{\tau}+\xi_0-\frac{\sigma}{B^{1/4}}-k_1\frac{\hat{\tau}^2}{2B^{1/2}})^2.$$
Then, we formally expand this operator in powers of $B$ and, for $|\sigma|\leq MB^{\eta}$, with $\eta'>0$ small enough :
$$\got{h}(\sigma,B)=\got{h}_0+B^{-1/4}\got{h}_1+B^{-1/2}\got{h}_2+O(B^{-3/4+3\eta}),$$
where 
$$\got{h}_0=-\frac{d^2}{d\hat{\tau}^2}+(\hat{\tau}+\xi_0)^2,$$
$$\got{h}_1=-2(\hat{\tau}+\xi_0)\sigma,$$
$$\got{h}_2=k_0\hat{\tau}\frac{d}{d\hat{\tau}}-k_1\hat{\tau}^2(\hat{\tau}+\xi_0)+2k_0 \hat{\tau}(\hat{\tau}+\xi_0)^2+\sigma^2.$$
Thus, as in Section \ref{formal}, we compute a quasimode and obtain for some $\psi$ :
$$\|(\got{h}(\sigma,B)-(\la_0+\la_1 B^{-1/4}+\la_2 B^{-1/2}))\psi\|_{L^2(\R_+,(1-\frac{k_0\hat{\tau}}{B^{1/2}}))}=O(B^{-3/4+3\eta}).$$
Finally, we can prove that the previous operator admits only one eigenvalue strictly less than $1$ thanks to a comparison with the harmonic oscillator on a half axis and, applying the spectral theorem, we get the bottom of the spectrum given in (\ref{bottommodel}) (the values of $\sigma$ such that $|\sigma|\geq MB^{\eta}$ provide higher energies thanks to the non-degeneracy of $\xi\mapsto\mu(\xi)$ near $\xi_0$).
\paragraph{Return in the initial variables}~\\
Applying the Parseval formula, we get : 
\bg{align*}
&q_{mod}(u)=\widetilde{q_{mod}}(\tilde{v})\geq\\ &(\Theta_0+\Theta^{k_0,k_1}_{1/2}B^{-1/2})\int_{\underset{\hat{\tau}>0}{\hat{s}\in\R}}|v|^2\left(1-\frac{\hat{\tau}k_0}{B^{1/2}}\right)d\hat{\tau}d\hat{s}\\
&+B^{-1/2}\frac{\mu''(\xi_0)}{2}\int_{\underset{\hat{\tau}>0}{\hat{s}\in\R}}|D_{\hat{s}}\tilde{v}|^2\left(1-\frac{\hat{\tau}k_0}{B^{1/2}}\right)d\hat{\tau}d\hat{s}-CB^{-3/4+3\eta}\|u\|^2.
\end{align*}
We have : 
$$|D_{\hat{s}}\tilde{v}|^2=|(D_{\hat{s}}-\phi'(\hat{s}))v|^2.$$
As $|\phi'(\hat{s})|\leq C\hat{s}^2B^{-1/2}\leq CB^{-1/2+2\eta}$ on the support of $v$, we get :
$$|D_{\hat{s}}\tilde{v}|^2 \geq (1-B^{-1/4+\eta})|D_{\hat{s}}v|^2-B^{-1/4+\eta}|v|^2.$$
Moreover, we have : 
$$D_{\hat{s}}v=\frac{\alpha\hat{s}}{2B^{1/2}}f_B(s)^{-3}u+f_B(\hat{s})D_{\hat{s}}u.$$
We deduce : 
$$|D_{\hat{s}}\tilde{v}|^2\geq |D_{\hat{s}}u|^2-CB^{-1/4+\eta}(|u|^2+|D_{\hat{s}}u|^2)-B^{-1/4+\eta}|D_{\hat{s}}v|^2.$$
Recalling that 
$$d\hat{\tau}d\hat{s}=\left(1+\frac{\alpha\hat{s}}{B^{1/2}}\right)^{1/2}d\hat{t}d\hat{s},$$
$$|v|^2=\left(1+\frac{\alpha\hat{s}}{B^{1/2}}\right)^{1/2}|u|^2,$$
where $\hat{t}=B^{-1/2}t,$
and with the tangential Agmon estimates, we get :
$$\|D_{\hat{s}}v\|^2\leq C\|u\|^2,\quad \|D_{\hat{s}} u\|^2\leq C\|u\|^2$$
and
\bg{eqnarray}\label{lbmodel}
q_{mod}(u)&\geq& \Theta_0\|u\|^2+\Theta_{k_0,k_1}B^{-1/2}\|u\|^2\nonumber\\
&&+\left(\int \left\{\alpha\Theta_0 |\hat{s}\check{u}|^2+\frac{\mu''(\xi_0)}{2}|D_{\hat{s}} \check{u}|^2 d\hat{s}\right\}\left(1-\frac{\hat{t}k_0}{B^{1/2}}\right)d\hat{t} \right)B^{-1/2}\\
&&-CB^{-3/4+3\eta}\nonumber,
\end{eqnarray}
where $\check{u}(\hat{t},\hat{s})=u(\hat{\tau},\hat{s})$ and, thanks to the Agmon estimates, we have replaced $D_{\hat{s}}u$ by $D_{\hat{s}}\check{u}$ and $\hat{\tau}$ by $\hat{t}$ by noticing that $D_{\hat{s}}u=D_{\hat{s}}\check{u}+\frac{\dr\hat{t}}{\dr\hat{s}}D_{\hat{t}}\check{u}$ and $\la(\hat{s}B^{-1/4})\hat{\tau}=~\hat{t}$.
We recognize the quadratic form of the harmonic oscillator and we have :
$$\int \left\{\alpha\Theta_0 |\hat{s}\check{u}|^2+\frac{\mu''(\xi_0)}{2}|D_{\hat{s}} \check{u}|^2 d\hat{s}\right\}\geq\sqrt{\frac{\mu''(\xi_0)\alpha\Theta_0}{2}}\int |\check{u}|^2 d\hat{s}.$$
We take $\eta=\frac{1}{20}$ and the lower bound of Theorem \ref{preciseestimate} follows from (\ref{lbmodel}), (\ref{redminimum}) and (\ref{formulasBS}) after having noticed that the estimates of Agmon give :
$$\int_{\Om} |\chi_{j_{min}}u_B|^2 dx=(1+O(e^{-cB^{\eta}}))\int_{\Om} |u_B|^2 dx.$$
\section{Estimate for the third critical field of the Ginzburg-Landau functional}
In this section, we give an estimate of the third critical field of the Ginzburg-Landau functional in the case where the applied magnetic field denoted by $\beta$ admits a unique and non degenerate minimum on the boundary of $\Om$. The constant magnetic field case has already been studied in details (see \cite{FouHel4,Lupan, Lupan3, Lupan4}).
\paragraph{Recall of properties of the functional}~\\
The Ginzburg-Landau functional is defined by :
$$\mj{G}(\psi,\A)=\int_{\Om}\big\{|(i\nabla+\sigma\kappa\A)\psi|^2 -\kappa^2|\psi|^2+\frac{\kappa^2}{2}|\psi|^4\big\}dx+(\kappa\sigma)^2\int_{\Om}|\nabla\times\A-\beta|^2 dx,$$
for $\psi\in H^1(\Om,\C)$ and $\A\in H_{div}^1(\Om,\R^3)$ where 
$$H_{div}^1(\Om,\R^3)=\{\A\in H^1(\Om,\R^3)\,:\,\dive(\A)=0\,\mathrm{in}\,\Om,\A\cdot\nu=0\,\mathrm{on}\,\dr\Om\}.$$
We assume moreover that 
$$\beta=\nabla\times\F.$$
Then, we introduce the critical fields :
$$H_{C_3}(\kappa)=\inf\{\sigma>0\,:\, (0,\F)\mbox{ is the unique minimizer of }\mj{G}_{\kappa,\sigma}\},$$
$$\conj{H}_{C_3}(\kappa)=\inf\{\sigma>0\,:\, (0,\F)\mbox{ is the unique minimizer of}\,\mj{G}_{\kappa,\sigma' }\mbox{ for all }\sigma'>\sigma\},$$
$$\underline{H}_{C_3}(\kappa)=\inf\{\sigma>0\,:\, (0,\F)\mbox{ is a minimizer of}\,\mj{G}_{\kappa,\sigma}\}$$
and 
$$\conj{H}_{C_3}^{loc}(\kappa)=\sup\{\sigma>0\,|\,\la^1(\kappa\sigma\F)<\kappa^2\}.$$
We have
$$\underline{H}_{C_3}(\kappa)\leq H_{C_3}(\kappa)\leq \conj{H}_{C_3}(\kappa)$$
and
$$\conj{H}_{C_3}^{loc}(\kappa)\leq\conj{H}_{C_3}(\kappa).$$
We can prove the following result (cf. \cite{FouHel2}) :
\bg{theo}
Let $\Om$ be a bounded, simply connected domain with smooth boundary and suppose that the applied magnetic field $\beta$ satisfies
$$0<\Theta_0 b'<b.$$
Then, there exists $\kappa_0>0$ such that for all $\kappa\geq\kappa_0$ :
$$\conj{H}_{C_3}(\kappa)=\conj{H}_{C_3}^{loc}(\kappa).$$
Furthermore, if $B\mapsto\la^1(B\F)$ is strictly increasing for large $B$, then all the critical fields coincide for large $\kappa$ and are given by the unique solution $H$ of $$\la^1(\kappa H\F)=\kappa^2.$$
\end{theo}
\paragraph{Estimate of $H_{C_3}(\kappa)$ for large $\kappa$}~\\
Noticing that $B\mapsto\la^1(B\F)$ is strictly increasing for large $B$ (it is due to the exponential decrease of the first eigenfunctions away from the boundary, still true in the case of variable magnetic field ; see \cite[Chapter 9, Section 6]{FouHel2}), we deduce the following theorem :
\bg{theo}
Let $\Om$ be a bounded, simply connected domain with smooth boundary and suppose that the applied magnetic field $\beta$ has a unique and non degenerate minimum on $\dr\Om$ and that :
$$0<\Theta_0 b'<b.$$
Then, we have :
$$H_{C_3}(\kappa)=\frac{\kappa}{b'\Theta_0}-b'^{1/2}\frac{\Theta_{1/2}}{\Theta_0^{3/2}}+O(\kappa^{-7/20}).$$
\end{theo}
\paragraph{Acknowledgments}~\\
I am deeply grateful to Professor B. Helffer for his help, advice and comments. I would also like to thank A. Kachmar for his attentive reading and suggestions which improved the presentation of the paper.

\bibliographystyle{alpha}
\bibliography{biblio}
\end{document}